\documentclass[12pt]{article}%
\usepackage{amsmath}
\usepackage{amssymb}%
\usepackage{amsfonts}%
\usepackage{graphicx}
\input epsf.tex
\newcommand{\be}{\begin{equation}}
\newcommand{\ee}{\end{equation}}
\newcommand{\bea}{\begin{eqnarray}}
\newcommand{\eea}{\end{eqnarray}}

\begin{document}
\bigskip\begin{titlepage}
\begin{flushright}
UUITP-23/05\\
hep-th/0511273
\end{flushright}
\vspace{1cm}
\begin{center}
{\Large\bf Inflation as a probe of new physics\\}
\end{center}
\vspace{3mm}
\begin{center}
{\large
Ulf H.\ Danielsson} \\
\vspace{5mm}
Institutionen f\"or Teoretisk Fysik, Uppsala Universitet, \\
Box 803, SE-751 08
Uppsala, Sweden
\vspace{5mm}
{\tt
ulf@teorfys.uu.se \\
}
\end{center}
\vspace{5mm}
\begin{center}
{\large \bf Abstract}
\end{center}
\noindent
In this paper we consider inflation as a probe of new physics near the string or Planck scale. We discuss how
new physics can be captured by the choice of vacuum, and how this leads to modifications of the primordial spectrum as well
as the way in which the universe expands during inflation. Provided there is a large number of fields contributing to the
vacuum energy -- as typically is expected in string theory -- we will argue that both types of effects can be present
simultaneously and be of observational relevance.  Our conclusion is that the ambiguity in choice of vacuum is an interesting
new parameter in serious model building.
\vfill
\begin{flushleft}
November 2005
\end{flushleft}
\end{titlepage}\newpage


\section{Introduction}

\bigskip

Is there a simple and unique theory describing the origin of the universe? To
some extent the answer is essentially yes. The key is inflation, which
provides a model for the initial conditions of the universe largely
insensitive to the details of physics near the Planck scale. Whatever happened
at the very first instance, the universe would, according to inflation, end up
roughly in the same state as the one in which we find it today. An excellent
introduction to the subject of inflation with references can be found in
\cite{Linde:2005ht}. Nevertheless, it is reasonable to look further and find
the reasons behind inflation and why it took place. String theory has recently
reached a sufficient state of maturity to be able to address these issues and
an intriguing new scenario is developing. Far from starting off in a unique
state, an enormous range of possibilities has been discovered. Some can be
described by known building blocks like branes and fluxes, but one can be sure
that there are still others to be revealed. Many authors take this as a sign
that chance played an important role in selecting our universe. The conclusion
-- for some depressing and for others stimulating -- seems to be that the
universe is, and always has been, a messy place.

Most attempts to describe the early universe use classical effective theory
where quantum effects never play a prominent role. Recent notable exceptions,
where true quantum cosmology is considered, can be found in
e.g.\cite{Firouzjahi:2004mx}\cite{Sarangi:2005cs}\cite{Brustein:2005yn}. The
reasons are obvious, these are the situations where reliable calculations can
be performed. But there is very little reason to believe that we will be able
to get all the way without the full force of string theory and quantum
gravity. In particular, having the above mentioned change in view of the
origin of the universe in mind where chance rather than simplicity is the
guide, it is natural to expect that whatever new effects we can imagine are
also likely to be present and of relevance if we look hard enough.

We will therefore consider a different line of approach where we temporarily
drop the hope of performing well controlled calculations. Instead we will look
for qualitatively new types of phenomena but limit ourselves to rough
estimates of their magnitude. The objective is to find examples where the
estimates suggest signatures within observational reach. In doing so we will
be pushing the limits of how insensitive inflation really is to unknown
physics at higher energies or earlier times. To do this we assume that all
effects due to unknown high energy quantum gravity or stringy physics can be
captured by a choice of vacuum. This assumption is extremely natural in an
expanding universe as has been spelled out in \cite{Danielsson:2002kx}. To
understand why, consider, for simplicity, a massless field with modes that
redshift as the universe expands. The key to the argument is that any given
field mode can not reliably be traced further back in time than to an era when
the wavelength was of order the fundamental length scale. At this point, as
argued in \cite{Danielsson:2002kx}, we parametrize our ignorance of what took
place at even earlier times and smaller length scales through the choice of
vacuum. The picture we have in mind is one where the mode emerges out of the
space time foam in a particular state determined by high energy physics. All
our calculations, however, will take place at low energies with crucial
initial conditions imposed not at a fixed time but at a fixed scale.

As argued in \cite{Danielsson:2002kx} and \cite{Danielsson:2002qh} there is a
natural expectation on the range of vacua that can be expected and this leads
to definite predictions of the kind of effects one should look for. If we had
been allowed to follow a given mode arbitrarily far back in time we could have
argued for a unique vacuum, the Bunch-Davies vacuum, which also is the vacuum
typically used in calculations of the primordial spectrum. In the presence of
a fundamental scale there is an ambiguity characterized by the ratio of the
fundamental scale to the Hubble scale. It is this ambiguity that new physics
beyond the fundamental scale can exploit to influence physics at lower scales.

The first example on how this could happen is concerned with the possibility
of remaining traces of elusive high energy physics in the CMBR fluctuations or
in large scale structures. The basic idea is that the inflaton, whose quantum
fluctuations are responsible for the fluctuations in the CMBR, is probing
physics near the string or Planck scale through the choice of vacuum as
explained above. Arguments have been put forward that the natural size of
these effects are just on the border of what might be detectable by planned
CMBR-measurements, \cite{Bergstrom:2002yd}. A positive detection would open a
direct window to string theory and quantum gravity. The literature on the
subject is rich -- the list of references, [5-27], only include a selected few
of all the papers written on the subject.

A non-trivial vacuum choice opens up the issue of back reaction on the
geometry, which will be our second example on how new physics could lead to
observational signatures. Typically, the issue of quantum back reaction is
neglected and assumed to be part of the unsolved problem of the cosmological
constant. In the cases we will be considering the nontrivial dependence of the
vacuum energy on the cosmological parameters make the issue less clear cut. In
fact, the problem was early realized in \cite{Tanaka:2000jw}, where it was
correctly concluded that the effects in general were to small to cause any
problem. In \cite{Brandenberger:2004kx}\ and \cite{Danielsson:2004xw} it was
argued that far from being a problem, the presence of the vacuum contribution
is instead an intriguing possibility. The vacuum energy will not, it was
argued, in anyway spoil inflation but can under the right conditions
contribute in a constructive way to the creation of an inflationary phase.
There is even, as argued in \cite{Danielsson:2004xw}, an automatic slow roll
built into the model, which could be of observational interest.

In this paper we will investigate the possibility of a model that both has
detectable modulations in the CMBR-spectrum and quantum vacuum energy which
interferes with inflation. We begin by a short review of the essential results
for the amplitude of the modulation in section 2, and in section 3 we review
the issue of back reaction. In section 4 we put the pieces together and we end
by some conclusions and outlook.

\bigskip

\section{A modulated spectrum}

We begin by considering the effect on the CMBR, or the primordial spectrum in
general, due to new physics modelled by a choice of vacuum different from the
usual Bunch-Davies vacuum. Interestingly, the effects are quite generic.
According to the analysis of \ \cite{Danielsson:2002kx}, the typical effect to
be expected on the primordial spectrum is of the form%

\begin{equation}
P(k)=\left(  \frac{H}{\overset{\cdot}{\phi}}\right)  ^{2}\left(  \frac{H}%
{2\pi}\right)  ^{2}\left(  1-\frac{H}{\Lambda}\sin\left(  \frac{2\Lambda}%
{H}\right)  \right)  ,
\end{equation}
where we note a characteristic, relative amplitude of the correction given by
$\frac{H}{\Lambda}$, and a modulation sensitively dependent on how
$\frac{\Lambda}{H}$ changes with $k$. $\Lambda$ is the energy scale of the new
physics which could be the string scale or the Planck scale. The overall
factor in front of the expression is the well known amplitude for the
primordial spectrum due to fluctuations in the inflaton field. It is easy to
qualitatively understand the presence of the modulation. Just as there are
oscillations taking place after the fluctuations re-enter through the horizon,
giving rise to the acoustic peaks, there are oscillations taking place before
the fluctuations exit and freeze. These fluctuations are only present if we
make a non-trivial choice of state for the inflaton different from the
Bunch-Davies vacuum. The claim is that whatever the nature of the high energy
physics really is, a modulated spectrum of the above described form is to be expected.

In \cite{Bergstrom:2002yd} one can find a discussion of the phenomenological
relevance of this effect and how the magnitude is related to the
characteristic parameters describing the inflationary phase. Using the
standard slow roll approximation, assuming a single field inflaton with
potential $V\left(  \phi\right)  $, one has that the amplitude of the
primordial spectrum is given by
\begin{equation}
\left(  \frac{H}{\overset{\cdot}{\phi}}\right)  ^{2}\left(  \frac{H}{2\pi
}\right)  ^{2}=\frac{1}{24\pi^{2}M_{pl}^{4}}\frac{V}{\varepsilon},
\end{equation}
where
\begin{equation}
\varepsilon=\frac{M_{pl}^{2}}{2}\left(  \frac{V^{\prime}}{V}\right)  ^{2},
\end{equation}
is one of the slow roll parameters. According to measurements of the CMBR we
have
\begin{equation}
\frac{V^{1/4}}{\varepsilon^{1/4}}\sim0.027M_{pl},
\end{equation}
implying a relation between the Hubble constant and the slow roll parameter
according to
\begin{equation}
\frac{H}{M_{pl}}\sim4\cdot10^{-4}\sqrt{\varepsilon}.
\end{equation}
We now assume initial conditions imposed at $\Lambda=\gamma M_{pl}$, which
implies
\begin{equation}
\frac{\Delta k}{k}\sim\frac{\pi H}{\varepsilon\gamma M_{pl}}\sim\frac{\pi
\beta^{2}}{\sqrt{3}\gamma\sqrt{\varepsilon}}\sim1.3\cdot10^{-3}\frac{1}%
{\gamma\sqrt{\varepsilon}},
\end{equation}
and
\begin{equation}
\frac{H}{\Lambda}\sim4\cdot10^{-4}\frac{\sqrt{\varepsilon}}{\gamma
}.\label{eq:effect}%
\end{equation}
These latter two relations are what we need in order to produce our estimates.
In order to beat cosmic variance and get an effect which has a chance to be
detected we need $\frac{H}{\Lambda}\sim10^{-2}$, or, in other words,
$\frac{\sqrt{\varepsilon}}{\gamma}\sim20$. In order for the modulation to be
detectable in the CMBR we would furthermore like to have $\frac{\Delta k}%
{k}\sim\mathcal{O}\left(  1\right)  $ and therefore $\varepsilon\sim10^{-2}$.
Remarkably, this is perfectly consistent with what one would expect from a
generic heterotic string compactification with high string scale. The
conclusion in \cite{Bergstrom:2002yd}, that the effects may be detectable by
the upcoming Planck satellite, has essentially been verified in subsequent
work such as \cite{Easther:2004vq}. The latter represents the most complete
analysis to date.

But what if we are interested in a considerably lower string and Hubble scale?
This is highly relevant for many of the brane universe scenarios. We might
still arrange for a suitable value of the amplitude -- if both $H$ and the
string scale are small -- but with the accompanying small value of
$\varepsilon$, $\frac{\Delta k}{k}$ will in general be much too large for
observable modulations. Since the range of observationally relevant
e-foldings, when large scale structure is included, is about $10$, we can
accept $\frac{\Delta k}{k}\sim\mathcal{O}\left(  10\right)  $ but not much
more. In this case we might push down to $\varepsilon\sim10^{-3}$, but this
seems to be the limit.

On the other hand, with a sufficiently large $\frac{\Delta k}{k}$ we can raise
the amplitude much further without getting in conflict with the present
measurements of the CMBR. For instance, we can stick to $\varepsilon
\sim10^{-2}$ and consider an amplitude at the $10\%$ level keeping
$\frac{\Delta k}{k}\sim\mathcal{O}\left(  10\right)  $. This is an intriguing
possibility where we expect a non-trivial mixing between effects due to the
spectral parameter and effects due to the modulation. In case of an
observationally detected running of the spectral parameter, when comparing the
CMBR results with measurements of large scale structure, a modulation of the
form discussed here could be of relevance.

This is not the whole story, however, As we will see below there is an
interesting way to relax the constraints. We will return to this issue after
investigating how the non-trivial vacuum might backreact in the expansion.

\bigskip

\section{A modified expansion}

\bigskip

As explained in the introduction, the presence of a nontrivial vacuum,
motivated by the presence of unknown high energy physics, raises the issue of
backreaction. Clearly, this is related to the problem of the cosmological
constant and it is not obvious how this decouples from the issue of inflation
itself. How can a subtraction of quantum fluctuations be argued for? Focusing
on the contribution to the vacuum energy coming form the non-standard vacuum,
as compared with the Bunch-Davies vacuum, we find an additional energy density
given by%

\begin{equation}
\rho_{\Lambda}\sim\frac{1}{2\pi^{2}}\int_{0}^{\Lambda}dpp^{3}\frac{H^{2}%
}{\Lambda^{2}}=\frac{\Lambda^{2}H^{2}}{8\pi^{2}}.
\end{equation}
Clearly, this contribution to the vacuum energy can be neglected, to lowest
order, as long as $\Lambda\ll M_{p}$. This was the conclusion of
\cite{Tanaka:2000jw}. However, even if the effect is small it can affect the
slow roll parameters and the way the expansion of the universe changes. In
order to see this we must be a little bit more careful in our analysis.

Since $H$ will be changing with time, i.e. decrease, we must take this into
account when calculating the vacuum energy density. Modes with low momenta
were created at earlier times when the value of $H$ were larger, and there
will be an enhancement in the way these modes contribute to the energy
density. We therefore write%
\begin{equation}
\rho_{\Lambda}\left(  a\right)  =\frac{1}{2\pi^{2}}\int_{\varepsilon}%
^{\Lambda}dpp^{3}\frac{H^{2}\left(  \frac{ap}{\Lambda}\right)  }{\Lambda^{2}%
}=\frac{1}{2\pi^{2}}\frac{\Lambda^{2}}{a^{4}}\int_{a_{i}}^{a}dxx^{3}%
H^{2}\left(  x\right)  ,
\end{equation}
where we have introduced a low energy cutoff corresponding to the energy at
the time of observation of modes that started out at $\Lambda$ at some
arbitrary initial scale factor $a_{i}$.\footnote{Note that we are using
another convention for $\Lambda$ as compared to the paper
\cite{Danielsson:2004xw} in order to facilitate the comparison with the
expressions for the modulation.}

If we take a derivative of the energy density with respect to the scale factor
and use $\frac{d}{da}=\frac{1}{aH}\frac{d}{dt}$, we find%
\begin{equation}
\dot{\rho}_{\Lambda}+4H\rho_{\Lambda}=\frac{1}{2\pi^{2}}\Lambda^{2}%
H^{3},\label{eq:contQ}%
\end{equation}
and we conclude that we must include a source term. With additional matter
present, with an equation of state of the form%
\[
\dot{\rho}_{m}+3H\left(  \rho_{m}+p_{m}\right)  =0,
\]
it was found in \cite{Danielsson:2004xw} that the evolution is governed by%
\begin{equation}
\frac{d}{da}\left(  a^{5}HH^{\prime}\right)  =-\frac{8\Lambda^{2}}{3\pi
M_{p}^{2}}a^{3}H^{2}-\frac{4\pi}{M_{p}^{2}}a^{3}\left(  1+w_{m}\right)
\left(  1-3w_{m}\right)  \rho_{m},\label{gamrull}%
\end{equation}
where we let $%
\acute{}%
=\frac{d}{da}$. The above way of expressing the content of the Friedmann
equations might seem unfamiliar, but is really the natural way in the presence
of the source term.

There are two constants of integration in this framework, as indicated in the
solution given by%
\begin{equation}
H^{2}=C_{1}^{2}a^{-2n_{1}}+C_{2}^{2}a^{-2n_{2}}+\frac{8\pi}{3M_{p}^{2}}%
\frac{\left(  1+w_{m}\right)  \left(  1-3w_{m}\right)  }{\left(
1+w_{m}\right)  \left(  1-3w_{m}\right)  -\frac{16\Lambda^{2}}{9\pi M_{p}^{2}%
}}\rho_{m},\label{eq:h2c1c2}%
\end{equation}
where%
\begin{equation}
n_{1,2}=1\pm\sqrt{1-\frac{4\Lambda^{2}}{3\pi M_{p}^{2}}}.
\end{equation}
The first of the constants of integration corresponds to how much radiation
there is, the second one is an analogue of the cosmological constant. In
accordance with this we note the the last term in (\ref{gamrull}) and
(\ref{eq:h2c1c2}) vanishes in case of matter with an equation of state like
the one of radiation or a cosmological constant. The reason is that these
types of matter are already taken care of by the constants of integration, and
become part of the initial conditions. There is an interesting and important
difference from the sourceless case, though: the cosmological constant is
rolling. As argued in \cite{Danielsson:2004xw} this could have interesting
implications for the problem of the cosmological constant, but we will not
discuss this issue further in this paper.

In \cite{Danielsson:2004xw} it was argued that the modification of the
cosmological evolution due to the source did not spoil inflation. In fact,
without the matter contribution, it gives rise to a natural slow roll with a
slow roll parameter easily read off from (\ref{gamrull}) as%
\begin{equation}
\varepsilon=\frac{2\gamma^{2}}{3\pi},
\end{equation}
for small $\gamma=\frac{\Lambda}{M_{p}}$.

The analysis has so far dealt with a particularly natural example of how the
source term depends on the expansion of the universe. For completeness we can
generalize the above analysis to a more general dependence on the Hubble
constant. Assuming%
\begin{equation}
\rho_{\Lambda}\left(  a\right)  =\frac{1}{2\pi^{2}}\int_{\varepsilon}%
^{\Lambda}dpp^{3}g\left(  \frac{H\left(  \frac{ap}{\Lambda}\right)  }{\Lambda
}\right)  =\frac{1}{2\pi^{2}}\frac{\Lambda^{4}}{a^{4}}\int_{a_{i}}^{a}%
dxx^{3}g\left(  \frac{H\left(  x\right)  }{\Lambda}\right)
\end{equation}
where $g=h_{n}f^{n}$, $f=f\left(  z\right)  =\frac{H^{2}}{\Lambda^{2}}$, and
$z=a^{-4}$, we find%
\begin{equation}
f^{\prime\prime}=-\frac{\Lambda^{2}}{3\pi M_{p}^{2}}h_{n}\frac{1}{z^{2}}f^{n}.
\end{equation}
This is a differential equation of the form of Emden-Fowler. Apart from being
exactly solvable for $n=1$, as used above, one can also write down the
solution for $n=0$:%
\begin{equation}
H^{2}=C_{1}a^{-4}+C_{2}-\frac{4\Lambda^{2}}{3\pi M_{p}^{2}}h_{n}\ln a.
\end{equation}
Again we find a an effect which can be interpreted as a rolling cosmological constant.

The question that is our main concern in this paper is whether we can find a
non-trivial interplay between the vacuum energy driven expansion and the usual
inflaton. In particular we will see whether we can relax the constraints
considered in the previous section and find interesting effects also in brane
universe models.

\bigskip

\section{Putting things together}

\bigskip

Is there a way to combine the two effects in the previous sections so as to
make them both observationally interesting? We saw above that it was possible
to achieve a slow roll inflationary phase by using the backreaction from a
non-standard vacuum. A problem with the present simple model is, however, that
there is no natural way for inflation to end. For this we consider a model
where in addition to the quantum vacuum energy there also is an inflaton. The
inflaton takes the role of the matter field of the previous section, with a
slow roll version of its equation of motion given by%
\begin{equation}
3aH^{2}\phi^{\prime}=-\frac{dV}{d\phi}.\label{phirull}%
\end{equation}
The evolution equation for the Hubble constant now takes the form
\begin{equation}
\frac{d}{da}\left(  a^{5}HH^{\prime}\right)  =-\frac{8n\Lambda^{2}}{3\pi
M_{p}^{2}}a^{3}H^{2}-\frac{4\pi}{M_{p}^{2}}\frac{d}{da}\left(  a^{4}\left(
aH\phi^{\prime}\right)  ^{2}\right)  ,
\end{equation}
where we have generalized to an arbitrary number of fluctuating fields, $n$,
all in the non-standard vacuum. As pointed out in \cite{Starobinsky:2002rp},
it is reasonable to expect that all fields, not only the inflaton, is being
influenced by the unknown high energy physics in a similar way. We will, for
simplicity, assume that all of these fields are effectively massless.

There will in general be a complicated interplay between the two terms, but we
will focus on situations where one of them dominates. It will also be
important to figure out when one term takes over after the other. Let us
consider a couple of explicit examples. We start with a potential of the type
of chaotic inflation given by%

\begin{equation}
V=\frac{1}{2}m^{2}\phi^{2}.
\end{equation}
To this we should add the contribution from the vacuum energy, but since this
show up as a rolling cosmological constant through constants of integration,
we do not put it in explicitly here. We can easily solve (\ref{phirull}) with
the result%
\begin{equation}
\phi=\phi_{0}a^{-\frac{m^{2}}{3H^{2}}},
\end{equation}
and the evolution equation becomes%
\begin{equation}
\frac{d}{da}\left(  a^{5}HH^{\prime}\right)  =-\frac{8n\Lambda^{2}}{3\pi
M_{p}^{2}}a^{3}H^{2}-\frac{16\pi}{M_{p}^{2}}\phi^{2}\frac{m^{4}}{H^{4}}%
a^{3}H^{2}.
\end{equation}
The first term alone would suggest a slow roll parameter given by%
\[
\varepsilon=\frac{2n\gamma^{2}}{3\pi},
\]
while the second on its own would give%
\begin{equation}
\varepsilon_{\inf}=\frac{4\pi}{9}\frac{\phi^{2}}{M_{p}^{2}}\frac{m^{4}}{H^{4}%
}.\label{epsinfkaos}%
\end{equation}
We will be interested in a situation where the roll is dominated by the vacuum
energy, that is $\varepsilon\gg\varepsilon_{\inf}$, during the era when the
relevant fluctuations are leaving the horizon. As time proceeds, $H$ will
eventually come down \ far enough for $\varepsilon_{\inf}$ to dominate and
eventually put an end to inflation. Let us see what kind of restrictions we
can derive for the various parameters.

When it comes to the amplitude of the primordial spectrum the inflaton plays a
crucial role. It is the inflaton that determines when inflation is going to
end. The fluctuations calculated on a spatially flat surface are given by
$\delta\phi$, but on a comoving surface, determined by a fixed value of $\phi
$, one finds curvature perturbations given by $\mathcal{R}=\frac{H\delta\phi
}{\dot{\phi}}$, which remain constant after horizon exit. One might possibly
worry that this is no longer true since we have played around with the
conservation laws by introducing a source term. As argued in
\cite{Danielsson:2004xw}, however, the effect of the vacuum energy can be
mimicked by matter with a peculiar equation of state obeying the usual
conservation laws. In this way, we see that the overall amplitude as well as
the amplitude of the modulations are governed by $\varepsilon_{\inf}$ -- even
though $\varepsilon_{\inf}\ll\varepsilon$. Hence we have%
\begin{equation}
\frac{H}{\Lambda}\sim4\cdot10^{-4}\frac{\sqrt{\varepsilon_{\inf}}}{\gamma},
\end{equation}
and in case of the wave length of the modulation we find%
\begin{equation}
\frac{\Delta k}{k}\sim\frac{\pi H}{\varepsilon\gamma M_{pl}}\sim
1.3\cdot10^{-3}\frac{\sqrt{\varepsilon_{\inf}}}{\gamma\varepsilon},
\end{equation}
where also $\varepsilon$ appears. Interestingly, we now have a decoupling
between the constraints thanks to the different roles played by $\varepsilon
_{\inf}$ and $\varepsilon$. Arguing like in section 2 we find that these
requirements are satisfied if $\frac{\sqrt{\varepsilon_{\inf}}}{\gamma}\sim20$
(that is $\varepsilon_{\inf}\sim400\gamma^{2}$) and $\varepsilon\sim10^{-2}$.
We see that in order to have the effect of the quantum vacuum energy to be
comparable to the effect of the inflaton, we need a large number of fields,
$n\sim10^{3}$. If we want the quantum vacuum energy to dominate, we need $n$
to be even larger. Such numbers are nevertheless not unreasonable in GUT or
string theories. We furthermore note that the larger $n$ is, the smaller
Hubble constant and string scale is allowed.

After the era when the observationally relevant perturbations leave one can
typically allow for only up to $60$ e-foldings. The total number of
e-foldings, including those taking place earlier, can certainly be much
larger. It is crucial to check whether the inflaton is able to stop inflation
fast enough. In other words, will $\varepsilon_{\inf}$ increase fast enough?
Expressing the dependence\ on the scale factor in (\ref{epsinfkaos}) through
$a=e^{N}$, where $N$ is the number of e-foldings and putting $a=1$ when the
fluctuations leave the horizon, we find
\begin{equation}
\varepsilon_{\inf}\sim\frac{4\pi}{9}\frac{\phi_{0}^{2}}{M_{p}^{2}}\frac{m^{4}%
}{H_{0}^{4}}e^{4N\varepsilon-\frac{2m^{2}}{3H_{0}^{2}}Ne^{2N\varepsilon}}.
\end{equation}
We see that $\varepsilon_{\inf}$ starts off increasing with $N,$ reaches a
maximum and then decreases again. The crucial issue is whether the maximum
value is large enough and can be reached fast enough. Unfortunately, with
values as small as $N=60$ and $\varepsilon=10^{-2}$ we can not achieve an
$\varepsilon_{\inf}$ which starts off well below $\varepsilon$ and then
manages to take over simply because $e^{4N\varepsilon-\frac{2m^{2}}{3H_{0}%
^{2}}Ne^{2N\varepsilon}}<e^{4N\varepsilon}\lesssim10$. There is, nevertheless,
room for an interesting interplay with contributions of the same order.

We next turn to our second example where we assume a potential of the opposite
sign,%
\begin{equation}
V=-\frac{1}{2}m^{2}\phi^{2},\label{branpot}%
\end{equation}
where again we should remember that there also is a contribution from the
vacuum energy. The expression for $\varepsilon_{\inf}$ is the same in terms of
$\phi$, but we now have%
\begin{equation}
\phi=\phi_{0}a^{\frac{m^{2}}{3H^{2}}},
\end{equation}
with the inflaton rolling in the opposite direction, and%
\begin{equation}
\varepsilon_{\inf}\sim\frac{4\pi}{9}\frac{\phi_{0}^{2}}{M_{p}^{2}}\frac{m^{4}%
}{H_{0}^{4}}e^{4N\varepsilon+\frac{2m^{2}}{3H_{0}^{2}}Ne^{2N\varepsilon}}.
\end{equation}
The situation is now slightly better since we are guaranteed an end to
inflation if we just wait long enough. The numerics work out such that with
$\frac{m^{2}}{H_{0}^{2}}\sim0.02$ and $\phi_{0}$ a little less than the Planck
scale, inflation ends within the required 60 e-foldings.

Similar results can be obtained using other potentials, but these examples
suffice to show that there is room for interesting interplay between the two
sources of rolling, provided we have a large number of degrees of freedom
contributing to the vacuum energy. An important observation is that the
consistency relation for the size of the tensor contributions is violated in
these models. That is, even if the roll is not that slow, we still can have
the tensor perturbations heavily suppressed. The reason is the decoupling
between $\varepsilon_{\inf}$ which governs the roll of $\phi,$ and
$\varepsilon$ which governs the roll of $H$ (if $\varepsilon_{\inf}%
\ll\varepsilon$). This is a property common to multifield inflationary models
where the contribution from tensor modes also is suppressed. Interestingly,
things work out in a similar way for holographic inflation,
\cite{Verlinde:1999xm} (for a recent paper with references see
\cite{Kiritsis:2005bm}). In the case of holographic inflation there is also a
modification of the Friedmann equations leading to new behavior, and just as
in our model one needs a large number of fields to get an interesting effect,
\cite{Lidsey:2005nt}.

\section{Conclusions and speculations}

\bigskip

We have seen how new physics captured by a nontrivial vacuum choice can lead
to generic effects on the primordial spectrum and the expansion of the early
universe. There is even a possibility of having both effects present provided
there is a large number of fields contributing to the quantum vacuum energy.
But how can we embed this phenomenological approach within string theory? How
do we make real calculations?

In a realistic scenario we would expect corrections to the potential and it is
reasonable to let those determine the physics of the end of inflation. In
fact, the models used for brane inflation can more or less be taken over
directly. In models such as those discussed in \cite{Kachru:2003sx} and
\cite{Iizuka:2004ct} the potential takes the form (\ref{branpot}) for small
values of $\phi$, but then rapidly gets corrections as the branes move closer.
In this way it is easy to get an $\varepsilon_{\inf}$ which dominates within
the required 60 e-foldings. It is far from clear, however, how to obtain
explicit constraints on the various parameters. How is the non-trivial vacua
to be understood from the higher dimensional string theory point of view? In
the brane universe scenarios the main contribution to the vacuum energy
driving inflation comes from the tension of the branes, adjusted by warp
factors. How is this changed in the presence of the rolling due to vacuum
fluctuations? The difficulties are a reflection on our inability to address
the problems of quantum gravity in a realistic setting. The answers should,
clearly, be given by string theory and it would be interesting to incorporate
the physics discussed in the present paper into a full-fledged brane universe model.

The effects we have studied influence the form of the primordial spectrum and
interfere with the way the universe expands. We have argued that the magnitude
of the effects are such that they need to be taken into account in
inflationary model building. This suggests new possibilities for string theory
and quantum gravity to become observationally relevant.

\bigskip

\section*{Acknowledgments}

The author is a Royal Swedish Academy of Sciences Research Fellow supported by
a grant from the Knut and Alice Wallenberg Foundation. The work was also
supported by the Swedish Research Council (VR).

\end{document}